\documentclass[]{aa}  
\def\apj{ApJ}
\def\apjs{ApJS}   
\def\aap{A\&A}

\def\ueber#1#2{{\setbox0=\hbox{$#1$}%       
  \setbox1=\hbox to\wd0{\hss$ #2$\hss}% 
  \offinterlineskip 
  \vbox{\box1\box0}}{}}

\def\lesssim{\,\lower  1mm  \hbox{\ueber{\sim}{<}}\,}   
\def\grsim{\,\lower  1mm  \hbox{\ueber{\sim}{>}}\,}

\makeatletter

\let\@internalcite\cite                
\def\cite{\@ifstar{\citeyear}{\citefull}}
\def\citefull{\def\astroncite##1##2{##1##2}\@internalcite}
\def\citeyear{\def\astroncite##1##2{##2}\@internalcite}

\def\citeau{\def\astroncite##1##2{##1}\@internalcite}
\def\citen{\def\astroncite##1##2{##1 ##2)}\@internalcite}
\def\possesivcite{\def\astroncite##1##2{##1's (##2)}\@internalcite}

\def\@citex[#1]#2{\if@filesw\immediate\write\@auxout{\string\citation{#2}}\fi
  \def\@citea{}\@cite{\@for\@citeb:=#2\do  
    {\@citea\def\@citea{;  }\@ifundefined
       {b@\@citeb}{{\bf  ?}\@warning  
       {Citation  `\@citeb'  on page  \thepage  \space undefined}}%          
{\csname          b@\@citeb\endcsname}}}{#1}}          %

\def\@cite#1#2{#1\if@tempswa , #2\fi} 
\def\@biblabel#1{} \makeatother

\begin{document}

\title{On the  formation of Super-AGB  stars in  intermediate  mass close binary
systems}

\author{Pilar Gil--Pons\inst{1} \and Enrique Garc\'{\i}a--Berro\inst{1,2}}
\titlerunning{Formation of Super-AGB stars in IMCBS}
\authorrunning{P. Gil--Pons \& E. Garc\'\i a--Berro}
	   
\institute{$^1$Departament   de   F\'\i   sica   Aplicada,   Universitat
	       Polit\`ecnica  de Catalunya,  c/Jordi Girona s/n, M\'odul
	       B-4,  Campus  Nord,  08034  Barcelona,   Spain,  (e-mail:
	       pilar,   garcia@fa.upc.es)\\   
	   $^2$Institute  for   Space   Studies  of  Catalonia,   c/Gran
	       Capit\'a 2--4, Edif. Nexus 104, 08034 Barcelona, Spain}

\date{Received 25 March 2002/Accepted 10 September 2002}

\abstract{  
The evolution of a star of initial mass 9 $M_{\odot}$,  and  metallicity
$Z = 0.02$ in a Close Binary System (CBS) is followed in the presence of
different mass companions in order to study their influence on the final
evolutionary stages and, in particular, on the structure and composition
of the  remnant  components.  In order to do that, we study two  extreme
cases.  In the first  one the mass of the  secondary  is 8  $M_{\odot}$,
whereas in the second one the mass was assumed to be 1 $M_{\odot}$.  For
the first of those cases we have also explored the possible  outcomes of
both conservative and  non-conservative  mass-loss episodes.  During the
first mass  transfer  episode,  several  differences  arise  between the
models.  The system with the more extreme  mass ratio  ($q=0.1$)  is not
able to survive the first Roche lobe  overflow  (RLOF) as a binary,  but
instead,  spiral-in of the  secondary  onto the  envelope of the primary
star is most likely.  The system formed by two stars of comparable  mass
undergoes two mass transfer  episodes in which the primary is the donor.
We have performed two sets of calculations corresponding to this case in
order to account for  conservative  and  non-conservative  mass transfer
during the first mass loss episode.  One of our main results is that for
the  non-conservative  case the secondary becomes a Super--AGB star in a
binary system.  Such a star undergoes a final dredge-up episode, similar
to that of a single star of comparable mass.  The primary  components do
not undergo a Super--AGB phase, but instead a carbon-oxygen  white dwarf
is formed in both  cases  (conservative  and  non-conservative),  before
reversal mass transfer  occurs.  However,  given the extreme mass ratios
at this stage between the  components of the binary  system,  especially
for the  conservative  case, the possibility of merger episodes  remains
likely.  We also discuss the presumable final outcomes of the system and
possible observational counterparts.
 }

\maketitle

%_______________________________________________________________________

\section{Introduction}

This is the  second of a series of papers in which we intend to  explore
extensively  the  evolution of  intermediate  mass close binary  systems
(IMCBS).  IMCBS  are  defined  as those  systems  in which  the  primary
component  develops a partially  degenerate  carbon-oxygen  core,  after
burning central helium in non-degenerate  conditions.  In particular, we
focus on the evolution of  heavy-weight  intermediate  mass stars, which
have primary masses between $\sim\,8\,M_\odot$ and $11\,M_\odot$.  These
stars are thought to  ultimately  develop ONe cores  (Ritossa,  Garc\'\i
a--Berro  \& Iben  1996), and  populate  the  brightest  portion  of the
Asymptotic  Giant Branch.  In this paper we follow the evolution of both
components  of the binary  system from their main  sequence  phase until
late evolutionary stages, paying special attention to the carbon burning
phase, and to the final objects we can encounter afterwards.

Even  though the  evolution  of  isolated  intermediate  mass  stars has
recently been analyzed --- see, for instance, Ritossa, Garc\'\i a--Berro
\& Iben (1999), and references therein --- the evolution of this kind of
star in binary  systems has been very little  studied.  Probably  one of
the reasons for the absence of this kind of model in the  literature  is
that the conversion of a CO core into an ONe core involves following the
evolution of an unstable  nuclear  burning flame as it propagates  first
towards the center of the star and then outwards.  The  calculation is a
very delicate and time-consuming task since the propagation of the flame
front is not steady, but is  interrupted  by a series of rather  violent
shell flashes  (Garc\'\i  a--Berro,  Ritossa \& Iben 1997).  The flashes
are violent  because the  nuclear  reaction  rates are very  temperature
sensitive and the material is partially  degenerate.  Moreover,  because
of the near discontinuity in the physical  variables at the flame front,
following  the inward  motion of the flame  requires  very good  spatial
resolution  and short  time-steps.  Furthermore,  the evolution of stars
within this range of masses in binary systems is even more  complicated,
since the  presence  of a close  companion  can  dramatically  alter the
evolution  of the primary  star and its final  outcome.  As a result the
only recent  calculation in which a heavy-weight  intermediate mass star
is followed  from its main  sequence  phase until carbon is exhausted in
the core is that of Gil--Pons \& Garc\'\i  a--Berro (2001), who followed
the evolution of a $10\,M_\odot$ star with solar metallicity in a binary
system.  Moreover,  most of the  developments  in the field of  binaries
frequently  disregarded  the study of AGB stars, arguing that this phase
is prematurely  quenched in binary systems, due to significant mass loss
in the evolution  previous to the AGB phase.  Important  exceptions  are
the works of Jorissen  (1999) and Smith et al.  (1996) who  proposed the
existence of AGB stars in binary systems and, more  recently, Van Eck et
al.  (2001),  who found  observational  evidence  of three AGB  stars in
binary systems.

In our  previous  paper  (Gil--Pons  \&  Garc\'{\i}a--Berro,  2001),  we
proposed a scenario  in which a $10\,  M_{\odot}$  star could  evolve to
become a Super-AGB star and, ultimately, an ONe white dwarf, in spite of
the fact  that all the  hydrogen-rich  envelope  and most of the  helium
layer had been lost in previous  phases of the evolution.  In this paper
we consider again such an scenario but we relax some of the  simplifying
assumptions  that  were  made  there.  In  particular,  we  compute  the
evolution of a $9\, M_{\odot}$ primary (the initially most massive star)
of the  binary  system,  and we  explore  the  effect of the mass of its
companion.  In doing so, we study two  extreme  cases; in the  first one
the initial mass ratio ($q_0$) is close to one (scenario  1), whereas in
our second  calculation  this ratio is less than 0.2  (scenario  2).  In
fact, $q_0=0.2$ is a theoretical limit under which spiral-in of the less
massive  star  onto  the  most  massive  and  finally,   the  merger  of
components, cannot be avoided --- see, e.g., Vanbeveren et al.  (1998a).

Han et al.  (2000) did a  comprehensive  study of the evolution of IMCBS
in order to  determine  the  influence  of the  initial  mass  ratio and
orbital  period on the final  parameters  of close binary  systems whose
primary initial mass is in the range $1\, M_{\odot}$ to $8\, M_{\odot}$.
Although  they  only  consider  the  conservative  case,  a  significant
variation in the primary  remnant  masses is observed.  In this paper we
focus on more massive  primaries --- which have been little  studied ---
and we  extend  our  previous  calculations  in order  to  consider  the
evolution  of the  secondary.  Furthermore,  since  there are still many
uncertainties  in relation to whether mass transfer is  conservative  or
not, we have computed two  evolutionary  sequences  for the first of our
scenarios, and we have  considered  the two extreme cases that may occur
during the first mass loss episode:

\begin{itemize} 
 \item [] 1.a :  A CBS  composed of a $9\,  M_{\odot}$  star plus a $8 \,
		M_{\odot}$ companion.  The initial period is of about 10
		days, so that mass loss from the primary starts  shortly
		after  hydrogen  is ignited in a shell (a mass  transfer
		episode of the  $B_{\rm  r}$  type).  Mass  transfer  is
		conservative during the first RLOF.
\item []  1.b : A CBS  composed of a $9\,  M_{\odot}$  star plus a $8 \,
		M_{\odot}$  companion.  The  initial  period is of about
		150 days, so that mass loss from the primary starts when
		it is  climbing  the red giant  branch (a mass  transfer
		episode of the  $B_{\rm  c}$  type).  Mass  transfer  is
		non-conservative during the first RLOF.
\item []   2 :  A CBS  composed of a $9\,  M_{\odot}$  star plus a $1 \,
		M_{\odot}$  companion.  The initial period is of about 5
		days.
\end{itemize}

Case 1.a  corresponds to a typical  conservative  mass transfer  episode
(Nelson \&  Eggleton  2001; Han et al.  2000),  whereas  case 1.b can be
highly  non-conservative.  Therefore,  we  have  studied  it  under  the
assumption that a common envelope forms that removes an important amount
of mass and angular momentum from the system.  It is precisely from this
secondary of the non-conservative  case that a Super--AGB star develops.

The paper is  organized  as follows.  In section 2, we present in detail
our  evolutionary  scenarios.  Section 3 is devoted  to the study of the
evolution  of the  first of the  above  mentioned  scenarios  in which a
massive  companion is assumed, whereas in section 4 we explain in detail
the  evolution  in the  case  of a  low-mass  secondary.  The  phase  of
reversal mass transfer in the first of our scenarios is fully  explained
in section \S 5.  Finally in  Section 6 we  discuss  and  summarize  our
major findings.

%_______________________________________________________________________

\section{The scenarios}

A general overview of the evolutionary  scenarios  mentioned in \S 1 can
be found in Fig.  1, for the cases  1.a and 1.b, and in Fig.  2 for case
2.

\subsection{Cases 1.a and 1.b:  $9\, M_{\odot}$ + $8\, M_{\odot}$.}

\begin{figure}[t]   
\vspace{7.9cm}   
\hspace{-2.7cm}   
\includegraphics{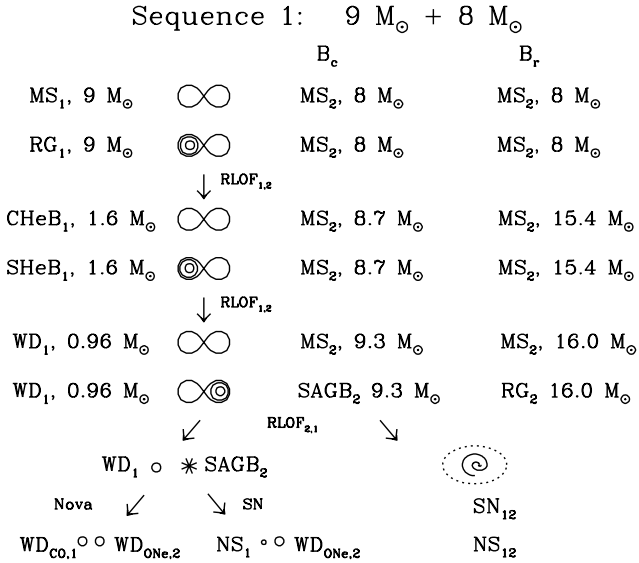}  
\caption{General  overview  of the  evolution  of the binary  system and
	 outline of the possible  final  outcomes, for the cases 1.a and
	 1.b,  see  text   for   details.  CHeB   stands   for   central
	 He--burning,  SHeB for shell  He--burning,  SAGB for Super--AGB
	 and the rest of the acronyms have their usual meaning.}
\end{figure}

The cases 1.a and 1.b (see Fig.  1) undergo two mass loss  episodes.  In
both  cases,  the  primary  remnant  has lost most of its  hydrogen-rich
envelope  after the first  one, but for the case 1.a, this mass is $\sim
1.8\, M_{\odot}$,  slightly higher than the mass of the case 1.b remnant
($\sim  1.6\,  M_{\odot}$).  Depending  on  whether  the  mass  transfer
process is conservative  or not the final mass of the secondary  changes
accordingly, with the  corresponding  values 8.7 and $15.4\,  M_{\odot}$
for the  $B_{\rm  c}$ and  $B_{\rm  r}$ cases,  respectively.  The final
orbital  periods  also change  after this phase, to $200$ and $120$ days
for   the   conservative   and   non-conservative   case   respectively.
Furthermore,  using the  prescription  of Iben \&  Tutukov,  (1985)  for
non-conservative  mass loss, we assume  that that the  angular  momentum
losses  during  the  common  envelope  phase  are as high as  70-80  \%.
Orbital shrinkage is a direct consequence of this kind of mass loss, and
therefore  one can expect  that the orbit of the binary  system  will be
wider for the case of conservative  mass transfer.  We will come back to
this point in \S 3.2.  Despite these differences, the next mass transfer
episode  for both cases  takes  place  during the  helium-shell  burning
phase.  At the end of the second mass  transfer  episode, the remnant of
the  primary  consists  of a  carbon-oxygen  core  of  about  $\sim  1\,
M_{\odot}$  surrounded  by  a  helium  envelope  of  about  $\sim  0.6\,
M_\odot$.  We  would  like  to  emphasize  that  considering  convective
overshooting  might have led to somewhat  more massive  remnants for the
primary.  However,  as we intend to keep  consistent  with our  previous
paper, and because no  significant  differences  are  expected,  we have
adopted the standard Schwarzschild criterion with no overshoot.

At the end of the second  mass  transfer  processes,  the  masses of the
companions are 9.3 and $16.0\,  M_{\odot}$  and the orbital  periods are
$1000$ and $700$ days for cases 1.a and 1.b, respectively.  After a very
short phase in which carbon is partially burnt off-center in the core of
the primary, the bulk of the carbon burning phase is ultimately avoided,
and the remnant of the primary star is a CO white  dwarf.  At some point
during the white dwarf cooling phase, the secondary  becomes a red giant
and reversal mass  transfer  ensues.  The final  outcome  depends on the
mass  ratio of the  binary  system.  Given  that at this  point the mass
ratio is $q\leq 0.2$,  spiral-in  of the white dwarf onto the  secondary
and,  consequently,  a merger of the two stars cannot be discarded.  If,
otherwise,  the system is able to survive in spite of the  extreme  mass
ratio,  one  can  expect  several  different  outcomes,  which  will  be
discussed in detail in \S 5.

\begin{figure}[t]   
\vspace{7.9cm}   
\hspace{-2.7cm}   
\includegraphics{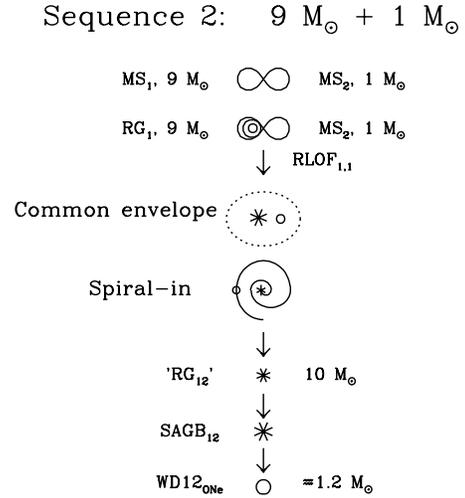}  
\caption{General overview of the evolution of the binary system, for the
	 case 2 --- see text for details.}
\end{figure}

\subsection{Case 2: $9\,M_{\odot}$ + $1\,M_{\odot}$.}

Fig.  2 shows  schematically the main  evolutionary  stages of our third
evolutionary  scenario.  For this case there is only one episode of mass
transfer,  occuring when the primary reaches red giant  dimensions.  The
most likely outcome is spiral-in of the secondary  onto the primary and,
ultimately,   the  merging  of  both   components.  This  happens  as  a
consequence  of their  initial mass ratio ($q_0 \leq 0.2$).  We simulate
the process as a thermohaline  mixing --- see, for instance,  Vanbeveren
et al.  (1998b).  That is, we assume that all the material  belonging to
the  secondary  mixes   instantaneously   and  homogeneously   with  the
hydrogen-rich envelope of the primary.  The final result of this process
is a $10\, M_{\odot}$  star.  However, this star will have the core of a
$9\,  M_{\odot}$   star.  A  consequence  of  the  merger  is  that  the
composition  of the  envelope  is  slightly  changed,  showing  a slight
overabuncances of C and O and an underabundance of N.  After the merger,
the product  continues its  evolution as a single star, very much in the
way described by Garc\'{\i}a-Berro \& Iben (1994).

%_______________________________________________________________________

\section{Evolution  of the 9 $M_{\odot}$ and 8 $M_{\odot}$  system until
	 reversal mass transfer}

\subsection{Evolution in the Hertzsprung-Russell diagram}

\begin{figure}    
\vspace{7.9cm}    
\hspace{-2.7cm}    
\includegraphics{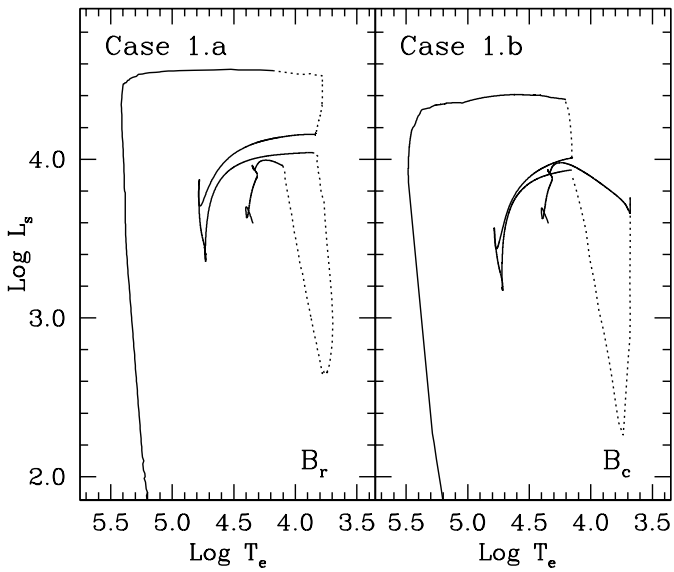}  
\caption[]{Evolutionary  tracks of the primary  components  of cases 1.a
	   and 1.b in the  Hertzsprung--Russell  diagram.  Mass transfer
	   episodes are shown as dotted lines.}
\end{figure}

In Fig.  3 the evolutionary tracks in the Hertzsprung-Russell diagram of
the primary  components  of the CBS for both the case 1.a --- left panel
--- and for the case 1.b --- right  panel --- are  shown.  The mass loss
episodes are  represented  as dotted  lines.  The primary  starts losing
mass just after the hydrogen  core burning  phase for case 1.a,  whereas
for the case 1.b the primary  begins its RLOF when already  climbing the
red giant branch (RGB).  Once this process is completed  for both stars,
they make their way to the left of the  Hertzsprung-Russell  diagram and
become blue  stragglers,  since they are almost  completely  deprived of
their  hydrogen-rich  envelopes.  During  this  phase  the  bulk of core
He-burning  occurs, and once helium is depleted  in the inner  core, the
primaries evolve towards decreasing effective temperatures.

The second mass loss episode for both cases starts when helium  burns in
a shell.  At this  point the  luminosity  increases  at almost  constant
effective  temperature  for both cases.  The mass loss processes in both
cases  continue  while the  primaries  climb the giant branch and finish
when the  evolutionary  paths of these  stars are well  advanced  on the
horizontal  track  for  case  1.a,  and  just  at the  beginning  of the
horizontal track for case 1.b.  Finally, once the primary  components of
the CBS are deprived of most of their  envelope they become white dwarfs
and follow the  corresponding  cooling tracks.  On the contrary, the two
secondary  components (Fig.  4) look very different from each other.  In
particular,  the narrow loop feature  that  appears in both cases before
the top of the main  sequence is reached are a  consequence  of the very
high accretion  rates at which mass transfer  proceeds  during the first
RLOF.  The second phase of accretion is less apparent, as it takes place
at much smaller  accretion  rates.  Once both  episodes of mass transfer
are over, the secondary for case 1.b evolves rather regularly, following
the path on the  Hertzsprung-Russell  diagram of an isolated $\sim 9.3\,
M_{\odot}$, which is very similar to that of a $9\,M_\odot$ star studied
by Garc\'\i a--Berro et al.  (1997).

\begin{figure}[t]   
\vspace{7.9cm}   
\hspace{-2.7cm}   
\includegraphics{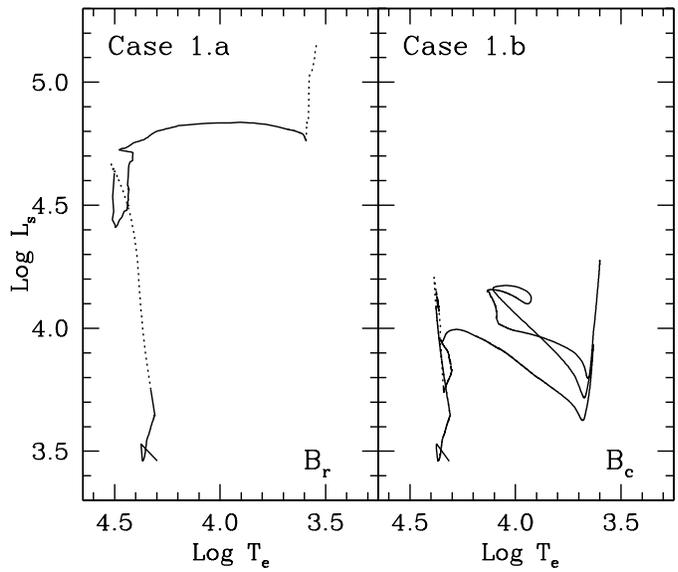}  
\caption[]{Evolutionary  tracks of the secondary components of cases 1.a
	   and 1.b in the  Hertzsprung--Russell.  Mass transfer episodes
	   are shown as dotted lines.}
\end{figure}

The evolution of the secondary  for the case 1.a requires a more careful
explanation.  It starts its evolution as a normal $8\, M_{\odot}$ single
star.  However, RLOF from its companion  drives a mass transfer  process
during its early core  hydrogen  burning  phase that allows it to almost
double  its  mass  (up to  $15.4\,  M_{\odot}$)  in a  relatively  brief
interval of time.  This  translates  into an  important  increase in the
luminosity (see the left panel of Fig.  4) and a slight  increase in the
effective temperature.  Accordingly, the surface radius of the secondary
increases  considerably  (from  $\sim  5\,R_\odot$  to about  $\sim  8\,
R_\odot$)  but not enough to overflow  its own Roche  lobe.  This effect
has been already noted by  Vanbeveren  et al.  (1998a) as being a direct
consequence  of  the  fact  that  the  thermodynamic  structure  of  the
secondary is not able to adapt to the rapidly accreted matter.

The second mass transfer  episode also happens  during the core hydrogen
burning  phase of the  secondary.  During  this  episode  the  amount of
accreted mass is relatively  modest  ($\simeq 1\,  M_{\odot}$),  and the
accretion rates are much lower.  The phase of core hydrogen  burning for
the secondary finishes shortly after this second accretion process.  The
subsequent  evolution in the  Hertzsprung-Russell  diagram is similar to
that of a high mass single star.

\subsection{The mass loss episodes.}

\begin{figure}[t]   
\vspace{7.9cm}   
\hspace{-2.7cm}   
\includegraphics{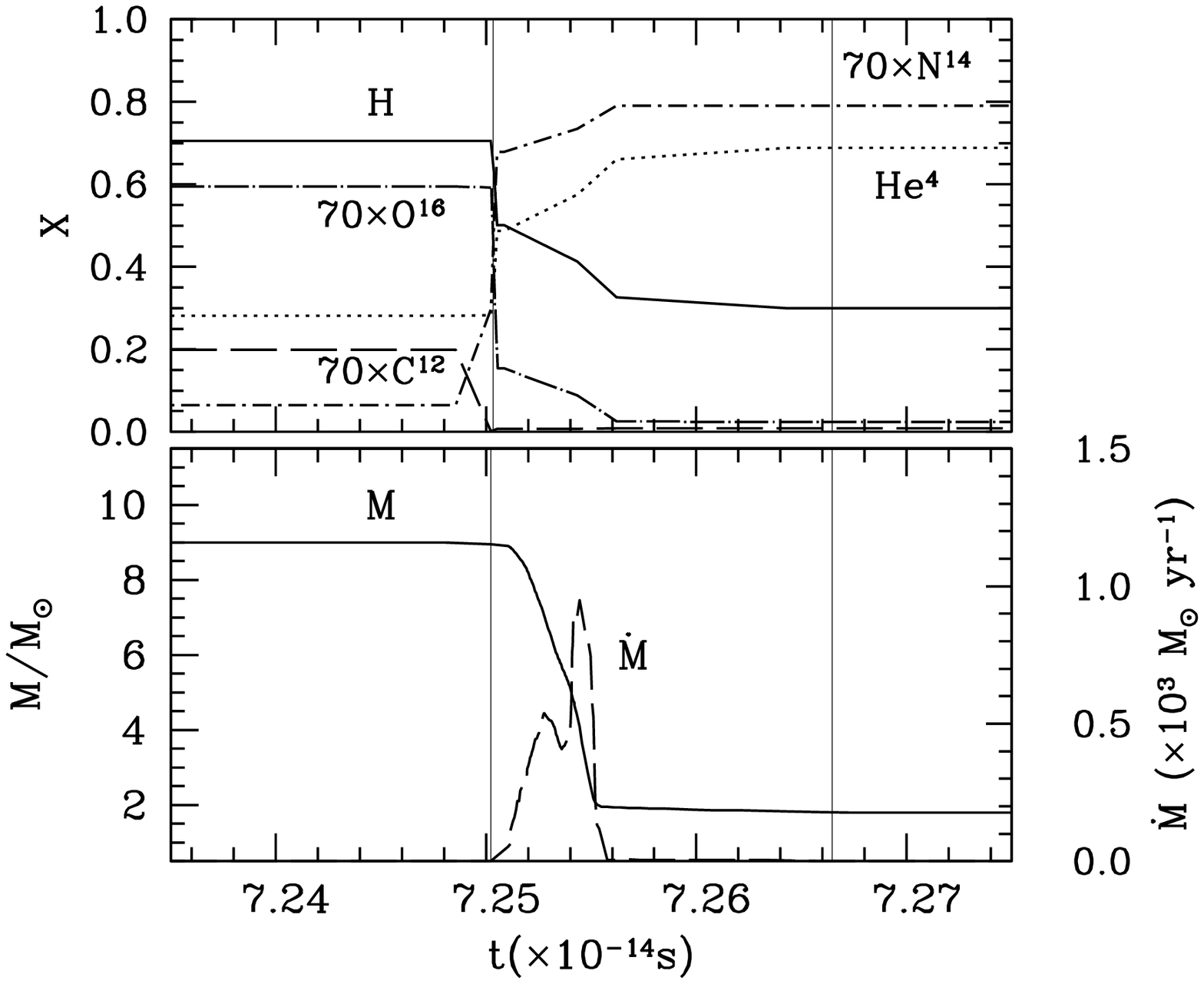}  
\caption[]{Upper  panel:  evolution  of the  surface  abundances  of the
	   primary for the case of a $B_{\rm r}$ mass transfer  episode.
	   Lower panel:  evolution of the mass and the mass loss rate of
	   the primary during the first RLOF episode.}
\end{figure}

As already mentioned, mass transfer in case 1.a begins shortly after the
primary burns hydrogen in a shell.  The presence of a radiative envelope
surrounding the mass donor allows for moderate mass transfer  rates.  In
fact, one can easily distinguish two phases in the mass loss episode ---
see the bottom  panel of Fig.  5.  The first  phase  corresponds  to the
beginning  of mass  transfer,  proceeds at thermal  time  scales and the
typical  mass loss rates are $\dot M \approx  10^{-4}  M_{\odot}\,  {\rm
yr}^{-1}$.  This phase  lasts for about  $10^4$  yr, and the bulk of the
hydrogen-rich  envelope is lost during it.  The second phase takes place
at much  lower  mass loss  rates  ($\dot M  \approx  10^{-6}  -  10^{-5}
M_{\odot}\, {\rm yr}^{-1}$), but it is considerably longer, lasting for,
approximately,  $4 \times  10^4$ yr.  The upper  panel of figure 5 shows
how the surface  abundances  change as the mass transfer  proceeds.  The
begining and the end of the mass transfer  episode are clearly marked by
the thin vertical lines.  As it can be seen in this panel, at the end of
the first mass loss episode the only CNO ash in the  remaining  envelope
is $^{14}$N.

For case 1.b the primary starts losing mass while it is climbing the red
giant branch.  Consequently, the presence of a deep convective  envelope
during this evolutionary stage induces mass loss to proceed on fast time
scales,  during a common  envelope  phase,  as was the case  studied  in
Gil--Pons \& Garc\'\i a--Berro (2001).  Since the first RLOF takes place
on very fast time scales, and since the surface radius of the primary is
well over the value of its Roche lobe radius during most of the time, we
let the mass  transfer  proceed in such a way that the Roche lobe radius
(Eggleton, 1983) is proportional to the surface radius.  As we intend to
study very  non-conservative  mass transfer episodes, we assume that the
Eddington   accretion   limit  is  valid,   and  we  do  not  allow  for
super--Eddington  accretion  rates, as was done in the  calculations  of
Podsialdowski et al.  (2001) and  Portegies--Zwart et al.  (2000).  That
is, we consider that all the matter lost by the primary at a rate higher
than the  Eddington  limit  is lost by the  system.  During  most of the
first RLOF  episode,  mass loss  proceeds at rates larger than  $10^{-3}
M_\odot\,  {\rm  yr}^{-1}$ and we obtain that only about $10 \% $ of the
mass lost by the primary is actually accreted by the secondary.

\begin{figure}[t]   
\vspace{7.9cm}   
\hspace{-2.7cm}   
\includegraphics{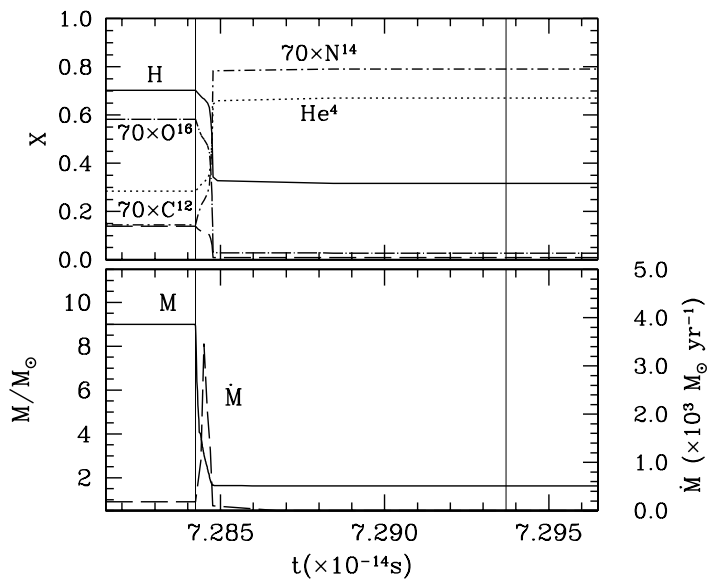}  
\caption[]{Upper  panel:  evolution  of the  surface  abundances  of the
	   primary for the case of a $B_{\rm c}$ mass transfer  episode.
	   Lower panel:  evolution of the mass and the mass loss rate of
	   the primary during the first RLOF episode.}
\end{figure}

As for the  conservative  case, for this one we can also  distinguish  a
fast and a slow phase in the mass  transfer  episode  --- see the bottom
panel of figure 6.  However,  in this case the fast  phase  yields  much
larger  mass  loss  rates  ($\dot M  \approx  10^{-3}  M_{\odot}\,  {\rm
yr}^{-1}$).  The whole RLOF episode  lasts for about $3 \times 10^4$ yr.
Again,  in the top  panel  of  figure  6 the  evolution  of the  surface
abundances as the mass transfer  episode  proceeds is shown.  It is easy
to see that the final composition at the surface is very similar in both
case 1.a and in case 1.b.  Our final  hydrogen  abundance  at the end of
the first mass loss episode  turns out to be $\simeq 0.3$ in both cases,
in very good agreement with the results of the  calculations of de Loore
\& Vanbeveren et al.  (1995).  The second mass loss episode  proceeds at
much lower transfer rates, and thus the secondary is able to accrete all
the matter that is lost by its companion,  with little  modification  of
its  thermodynamical  magnitudes  (see the right panels of figures 3 and
4).  This process starts after the development of a convective  envelope
surrounding  the primary as a consequence of helium shell burning and it
lasts until carbon is partially burnt in the helium-exhausted region.

\subsection{The structure and composition of the cores of the primary.}

Fig.  7 shows the abundance profile of the primary for our case 1.a, and
in Fig.  8 we show  those  abundance  profiles  for the  case  1.b.  The
differences of composition for the central regions are  negligible,  the
most  abundant  nuclei  being  oxygen  ($X_{\rm  O}$ = 0.62), and carbon
($X_{\rm C}$ = 0.37) in both cases.  Minor  differences in the shapes of
the chemical  profiles of the two cores appear as a  consequence  of the
peculiarities  of their  evolution,  specially in those  phases in which
mass loss or convection play an important role.  The main  difference is
the size of the  core,  that  happens  to be  smaller  for the  case 1.b
($M_{\rm CO} = 0.78\, M_{\odot}$), when compared to that of the case 1.a
($M_{\rm CO} = 0.93\,  M_{\odot}$).  This is a direct consequence of the
fact that, because the orbit is wider for case 1.a, the corresponding CO
core has more  time to grow  before a new RLOF  occurs.  The  similarity
between the primary  cores before  reversal  mass  transfer  for the two
cases   supports   one  of  the  main   hypotheses   in   Gil--Pons   \&
Garc\'{\i}a--Berro  (2001) at this stage of the  evolution,  namely that
the core of the primary is only slightly affected by the initial mass of
the secondary.

\begin{figure}    
\vspace{7.9cm}    
\hspace{-2.7cm}    
\includegraphics{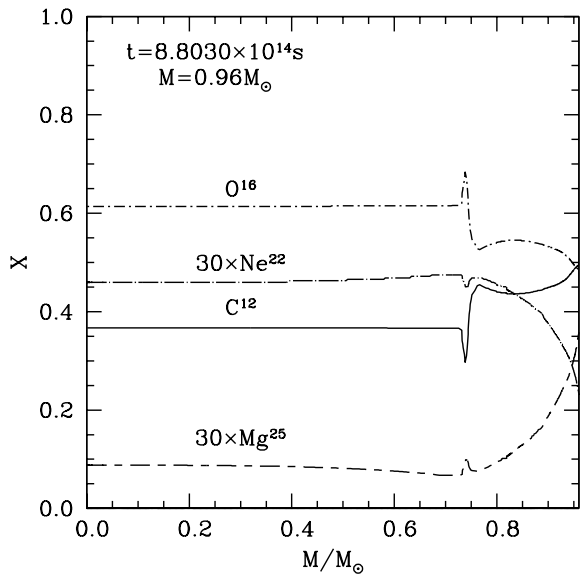} 
\caption[]{Abundance  profiles  of the  primary at the end of He burning
	   for case 1.a.}
\end{figure}

\subsection{The structure and composition of the secondary of Case 1.a}

The secondary for case 1.a undergoes two accretion  episodes while still
burning   hydrogen  in  its  core.  Computationally,   we  have  treated
accretion  in the usual way, the  so-called  standard  accretion  model.
That is, we have assumed that the infalling matter has the same specific
entropy as the surface layers of the mass-gaining component --- see, for
instance,  Pols (1994).  As noted by Vanbeveren  et al.  (1998a),  using
this approximation has two important consequences.  The first one is the
sudden increase in the luminosity of the  mass-gaining  component  which
has been already  commented on \S3.1.  The second one is the increase of
the  convective  core  as  accretion  occurs.  This  effect  is  clearly
illustrated  in the top  panel of Fig.  9.  The  times at which  the two
accretion  episodes begin are clearly marked by the thin vertical lines.
The total mass of the star is also shown as a thick  solid  line.  Since
the second episode occurs at a much more modest pace, this effect is not
as evident as it was for the first one.  On the other  hand, the  change
in the surface  composition  of the  secondary  is more  evident in this
case, as the accreted material onto its hydrogen-rich envelope is almost
pure  helium.  The change in the  composition  has the effect of driving
the evolution in the Hertzsprung-Russell  diagram to higher luminosities
than those  corresponding  to a star of the same mass and a non-polluted
envelope.

\begin{figure}[t]   
\vspace{7.9cm}   
\hspace{-2.7cm}   
\includegraphics{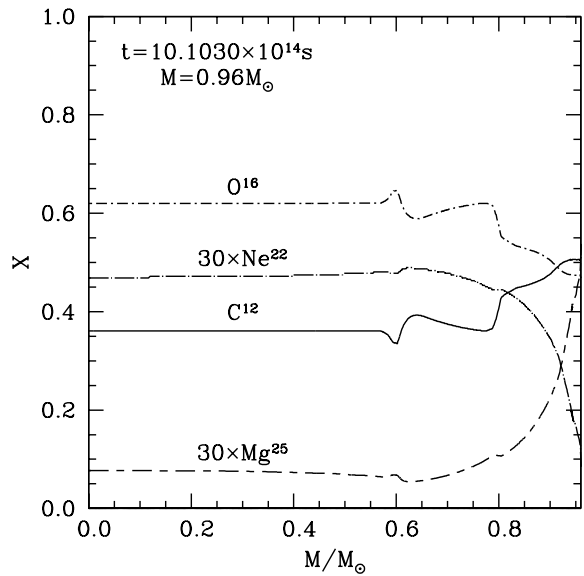} 
\caption[]{Abundance  profiles  of the  primary at the end of He burning
	   for case 1.b.}
\end{figure}

\begin{figure}[t]  
\vspace{7.9cm}    
\hspace{-2.7cm}    
\includegraphics{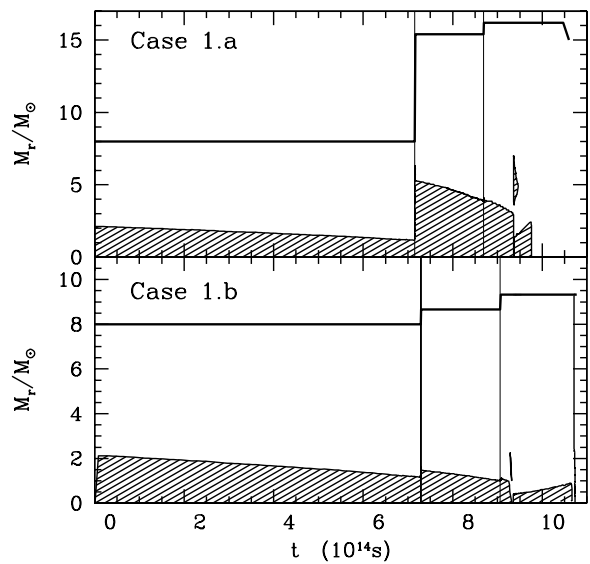}  
\caption[]{Convective  regions  for the  evolution  previous  to  carbon
	   burning, for the secondaries of both our case 1.a (top panel)
	   and 1.b (bottom panel).}
\end{figure}

\begin{figure*}[t]   
\vspace{9.4cm}   
\hspace{1.5cm}   
\includegraphics{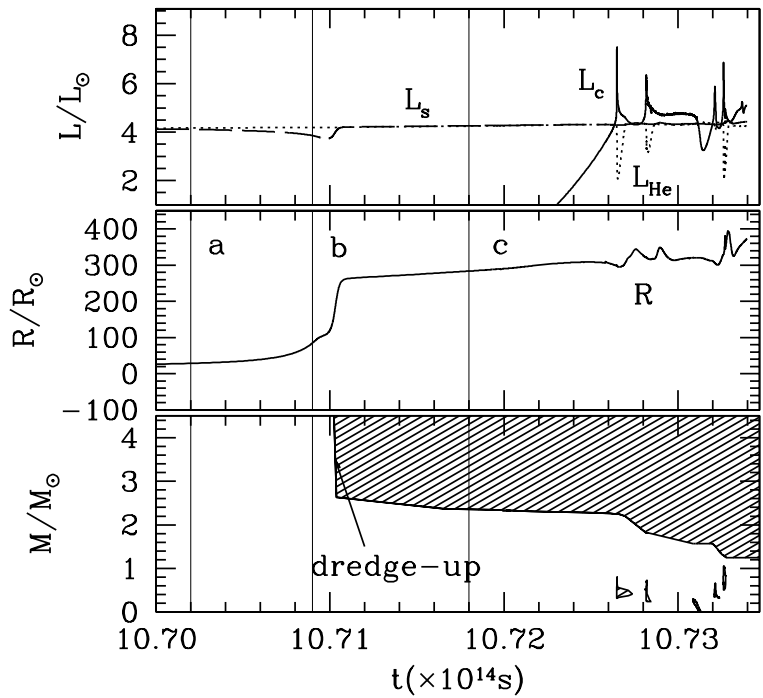}  
\caption[]{Evolution  of the main physical  parameters  of the secondary
	   during carbon burning for case 1.b.}
\end{figure*}

After these two accretion  episodes, the  secondary  evolves as a single
$16\, M_{\odot}$ star.  When the secondary reaches red giant dimensions,
superwinds   start   playing  an  important   role.  We  have  used  the
prescription  of Jura (1987) in order to take into account the mass-loss
rates due to stellar winds (see also Vanbeveren et al.  1998b).  For our
case the mass-loss rate turns out to be $\dot{M}_2 \sim 2\times  10^{-5}
M_\odot\, {\rm yr}^{-1}$.  The effect of mass loss from the secondary is
clearly  seen for long  enough  times  in the top  panel  of  figure  9.
However,  it should be taken into  account  that  several  uncertainties
surround this  evolutionary  phase, which can  significantly  effect the
evolution  of the binary  system  beyond this point.  The main source of
these  uncertainties  is the fact that the radius of the secondary  gets
very close to that of its Roche  lobe.  Therefore,  it is  difficult  to
ascertain  whether or not reversal mass  transfer  occurs.  Furthermore,
there is the possibility of enhancement of the wind mass-loss  rates due
to the presence of a close  companion or, even,  partial  capture of the
wind by the primary  remnant, as happens in symbiotic  systems  (Iben \&
Tutukov,  1996).  Moreover,  when carbon is ignited at the center of the
secondary,  this star has not yet  abandoned  the red giant  branch, and
both the helium and the hydrogen  burning shells remain active.  We have
not  followed  in detail  the  carbon  burning  phase  because  of these
uncertainties.  However, at this point of the  evolution  the  resultant
core is made of CO, with an oxygen abundance of $X_{\rm O} = 0.73$ and a
carbon   abundance   of   $X_{\rm   C}  =   0.23$.  The   size  of  this
helium--exhausted  core is $4.5\,  M_{\odot}$  and it is surrounded by a
helium envelope.  However, since the uncertainties in the wind mass-loss
rates are  significant, we  intentionally  refrain from giving a precise
value of its mass.

\subsection{The  structure and composition of the secondary of Case 1.b.
	    The evolution in the Super-AGB phase}

The evolution of the secondary component for case 1.b (non-conservative)
mass transfer presents some differences with respect to that of the case
1.a.  These  differences  are not important, as far as the two accretion
processes  are  concerned:  in fact, both of them occur  during the core
hydrogen--burning  phase of the mass gainer, and are  accompanied by the
same effects of overluminosity  and extra growth of the helium core (see
the top panel of figure 9).  Note however  that the first mass  transfer
episode  starts  earlier  for case 1.a than for the case  1.b, as can be
seen in figure  9.  The  differences  become  more  remarkable  once the
accretion  process is  finished.  The main reason for these  differences
is, of course, that the final masses of the secondaries are very unlike:
$16\,  M_{\odot}$ for the case 1.a, and $9.3\,  M_{\odot}$  for the case
1.b,  after  the two  accretion  episodes.  The  new  $9.3\,  M_{\odot}$
component  evolves as an  intermediate  mass star, very much in the same
way as the isolated $9\, M_{\odot}$  star studied by  Garc\'{\i}a--Berro
et al.  (1997).  In spite of the important  angular  momentum  loss that
the system undergoes  during the first mass transfer  process, the final
orbital  separation  after the second episode is such that the secondary
component  does not fill its Roche lobe until it reaches the  Super--AGB
phase.  In fact, this star is able to undergo a dredge-up  episode and a
series of thermonuclear flashes caused by carbon burning in a degenerate
core (see the top and bottom panels of Fig.  10).

\begin{figure}    
\vspace{7.9cm}    
\hspace{-2.7cm}    
\includegraphics{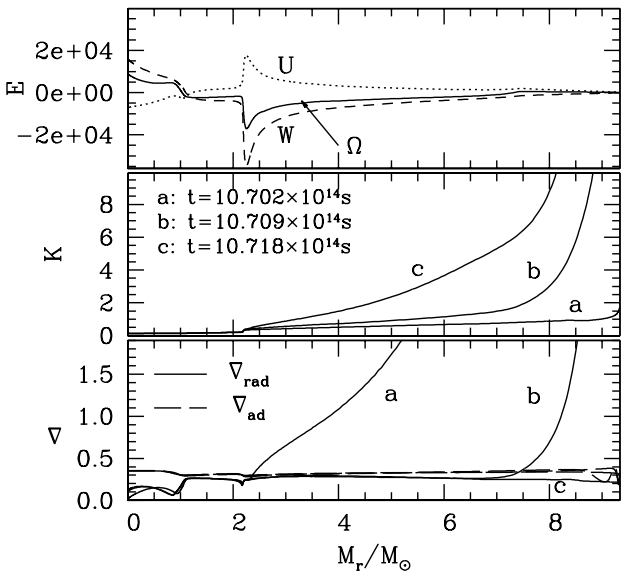}  
\caption[]{Energy  variation rates for model {\sl a} --- upper panel ---
	   and opacity and temperature gradient profiles for models {\sl
	   a},  {\sl  b} and  {\sl  c} ---  middle  and  bottom  panels,
	   respectively.}
\end{figure}

Opposite  to what  happens  to the  $9\,  M_{\odot}$  model  studied  by
Garc\'{\i}a--Berro  et al.  (1997),  this  dredge-up  episode  is  not a
consequence of energy  release  during the first carbon  burning  flash.
Instead,  it  is  the  gravothermal   energy  release  during  the  core
contraction phase which follows the exhaustion of central helium that is
responsible  for  the  expansion  and  cooling  of the  outer  envelope.
Figures  10 and 11 help  to  explain  this  phenomenon  by  showing  the
evolution of the most important model  parameters  during this dredge-up
episode.  In particular  the upper panel of Fig.  10 shows the evolution
of  carbon  and  helium  luminosity,   $L_{\rm  C}$  and  $L_{\rm  He}$,
respectively.  In this panel the  flashes  due to carbon  ignition  in a
degenerate  medium are  apparent  for times  longer than  $1.072  \times
10^{15}\,{\rm   s}$.  As  usual,   these   thermonuclear   flashes   are
accompanied  by the formation of  associated  convective  regions in the
inner  core (see the  bottom  panel of Fig.  10), and by  expansion  and
cooling of the upper layers, leading to the temporary  extinction of the
helium burning shell.  Note, however, that the dredge-up  episode is not
a consequence  of the carbon  flashes,  since these  flashes  occur well
after the fast  penetration of the convective  envelope, which occurs at
$t \simeq  1.071\times  10^{15}\, {\rm s}$.  Instead, in the lower panel
of  Figure  10 it can be seen  that the  time at  which  the fast  inner
penetration  of the base of the  convective  envelope is taking place is
practically  coincident  with  the  time at  which  the  surface  radius
increases at an almost constant surface luminosity, owing to an increase
of the core density.  In order to reinforce  this  argument, in Fig.  11
we show several relevant  physical  quantities for three selected models
at three  different  times:  before  the  fast  advance  inwards  of the
convective envelope (model {\sl a}), during the advance (model {\sl b}),
and after it (model {\sl c}).  The  location of these  models is clearly
shown in Fig.  10 and their corresponding times are labeled in Fig.  11.
The upper panel of figure 11 shows the terms involved in the equation of
energy  conservation  for model {\sl a}.  The  minimum in the solid line
corresponds to the gravitational energy release that, as can be seen, is
transformed  into work of  expansion  (dashed  line),  rather  than into
internal  energy  (dotted  line).  Note as well  that  this is done very
close  to the  inner  edge of the  contracting  envelope.  The  work  of
expansion  induces a considerable  increase in the surface radius of the
star  (central  panel of figure  10).  Ultimately,  the  cooling  of the
external layers has, as a main consequence, the increase in the envelope
opacity  ---  see  the  middle  panel  of  Fig.  11.  Consequently   the
radiative  gradient  increases  and,  at a  certain  point,  allows  the
convective envelope to move inwards.

\begin{figure}    
\vspace{7.9cm}    
\hspace{-2.7cm}    
\includegraphics{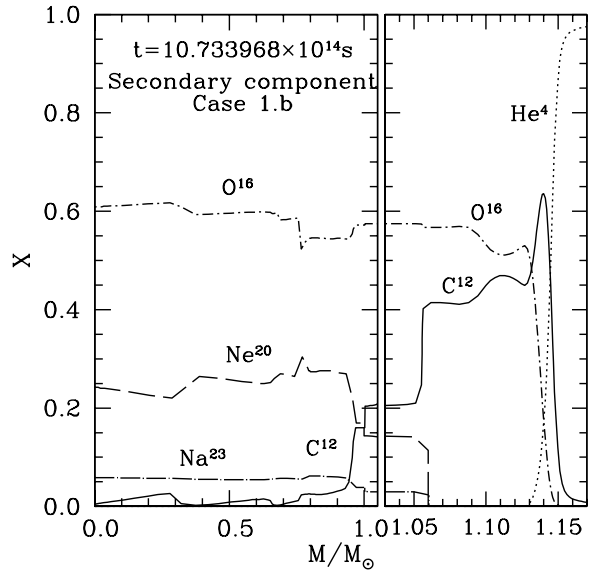} 
\caption[]{Abundance  profiles of the secondary  component at the end of
	   the calculation for case 1.b.}
\end{figure}

Finally, Fig.  12 shows the ONe core of the secondary  component for the
case 1.b.  The size of this core is of about $0.94\, M_{\odot}$, and the
main isotopes are oxygen  ($X_{\rm O} \simeq  0.60$), and neon  ($X_{\rm
Ne} \simeq 0.25$).  As can be seen in this figure, the neon-rich core is
surrounded  by a CO rich  layer of  irregular  profile,  as well as by a
helium rich  envelope, in  agreement  with the results of  Gil--Pons  \&
Garc\'\i a--Berro (2001).

\subsection{The reversal mass transfer}

At the end of the carbon burning phase the surface  radius has increased
considerably,  reaching  values close to the Roche lobe radius ($\sim \,
150 \, R_\odot$).  At this  evolutionary  stage the situation is similar
to that found in Gil--Pons  \&  Garc\'{\i}a--Berro  (2001).  Thus, it is
difficult to determine  the  mass-loss  rates, as there is the  combined
effect of RLOF and the typical superwinds of the AGB phase.

Perhaps the main  uncertainty in this phase of the evolution is the fact
that stellar winds from such a massive  secondary  might be enhanced ---
see, for  instance,  Iben \& Tutukov  (1996)  --- by the  presence  of a
companion.  Another  uncertainty  has to do with the  efficiency  of the
secondary  component in capturing the stellar wind.  In order to provide
an estimate of this  efficiency, we proceed as in Iben \& Tutukov (1996)
and we assume that the remnant of the primary is able to accrete all the
wind flowing into a disk that has the Bondi--Hoyle radius

\begin{equation} 
R_{\rm BH}=2\frac{GM_{\rm WD}}{v_{\rm w}} 
\end{equation}

\noindent  where $M_{\rm WD}$ is the mass of the remnant of the primary,
and $v_{\rm w}$ is the velocity of the wind flowing from the  secondary,
that we have taken to be about 25~km/s, which is a typical value for red
giants (Iben \& Tutukov 1996).  Note,  however, that this value can vary
significantly  for different red giant stars.  Under this assumption the
accretion rate is:

\begin{equation}    
\dot{M}_{\rm acc}=\dot{M}_{\rm GW}\frac{{R_{\rm BH}^2}}{4A^2}, 
\end{equation}

\noindent  where  $\dot{M}_{\rm  GW}$ is given by the expression of Jura
(1987), and $A$ is the  orbital  separation.  We get that the  accretion
rates are between $\sim 1\%$ and $\sim 5\%$ of the wind mass loss rates.
Therefore,  $\dot{M}_{\rm  acc} \sim 10^{-7} - 10^{-6}  M_{\odot}\, {\rm
yr}^{-1}$.  Analyzing  this mass  transfer  process is  certainly a very
delicate task, as the extreme  initial mass ratio between  components is
such that the possibility that the system would undergo a merger episode
is non negligible.  Hence, hydrodynamical  simulations should be carried
out in order to account for it in a realistic way --- see, for instance,
Hjellming \& Taam (1991), Iben \& Livio  (1993), Lai et al.  (1993) and,
more recently,  Taam \& Sandquist  (2000), and Ivanova \&  Podsiadlowski
(2001).  These  calculations  are beyond the scope of this  paper,  but,
still, we will make some  reasonable  estimates  of the  possible  final
outcomes in \S 5.

%_______________________________________________________________________

\section{Evolution of the $9\, M_{\odot}$ + $1\, M_{\odot}$ system.}

The system  composed of a $9\,  M_{\odot}$  and a $1\,  M_{\odot}$  star
evolves in the  following  way:  because the more  massive  star evolves
much faster than its  companion, it undergoes  RLOF while the  secondary
component   is  still   burning   hydrogen   in  its   core.  The   main
characteristic  that  determines the evolution of the system is, in this
case, the initial mass ratio between  components, $q = 0.1$.  This value
is smaller than the critical mass ratio and, thus, no matter  whether or
not the initial  period is such that the  primary  component  is able to
develop a deep  convective  envelope,  the system is forced to undergo a
merger episode  (Vanbeveren  et al.  1998a).  This merger occurs at time
$t = 7.3643 \times  10^{14}\,{\rm  s}$, and in order to follow it we use
the thermohaline  mixing  approximation.  Once the merger has ended, the
resulting single $10\,  M_{\odot}$ star behaves very much like the $10\,
M_{\odot}$ star studied by Garc\'{\i}a--Berro \& Iben (1994).

%_______________________________________________________________________

\section{Discussion and conclusions}

We have  followed  the  evolution  of a 9  $M_{\odot}$  star with  solar
metallicity  in close binary  systems  with  different  initial  orbital
parameters.  Our main goal has been to analyze the influence of the mass
of the secondary  component on the final possible  outcomes of the close
binary  systems.  In particular, we have computed a binary  evolutionary
scenario that allows the formation of  Super--AGB  stars in close binary
systems.  Our analysis encompassed all the relevant evolutionary phases,
starting from the main sequence of both components, following the helium
burning phase, and, if necessary the carbon burning  phase.  In summary,
we have studied the following three cases:

\begin{itemize}

\item [] Case 1.a:  9 $M_{\odot}$ + 8 $M_{\odot}$,  with initial orbital
	 period  $P_{\rm orb} \sim 10$ days, so the first mass  transfer
	 episode is a case $B_{\rm r}$ mass transfer episode.

\item [] Case 1.b:  9 $M_{\odot}$ + 8 $M_{\odot}$,  with initial orbital
	 period  $P_{\rm orb} \sim 150$ days, so the first mass transfer
	 episode is a case $B_{\rm c}$ mass transfer episode.

\item [] Case 2:  9 $M_{\odot}$ + 1  $M_{\odot}$,  with initial  orbital
	 period $P_{\rm orb} \sim 5$ days.

\end{itemize}

For the last of these three cases the first mass transfer  episode leads
to the merger of  components,  due to the  initial  mass  ratio  between
components:  $q \lesssim 0.2$.  The resulting  $10\,  M_{\odot}$  single
star behaves very much like the normal 10  $M_{\odot}$  star  studied by
Garc\'{\i}a--Berro and Iben (1994), once the merger episode is over.

Cases  1.a and 1.b  are  more  complicated.  These  two  binary  systems
undergo two mass  transfer  episodes.  This kind of  evolution  leads to
very  similar  primary  remnants:  both of them  are  massive  CO  white
dwarfs,  $M_{\rm WD} \sim 0.98\,  M_{\odot}$.  However,  the two systems
are very different in orbital period and masses of the secondaries.  For
case  1.a the  system  is  composed  of a  primary  remnant  and a $16\,
M_{\odot}$  main sequence  star, with a  hydrogen-rich  envelope  highly
polluted by the CNO products from the primary  remnant,  dredged--up  by
convection  and expelled in its second RLOF, and also by the products of
its own helium  burning  shell.  The secondary  for case 1.b is a $9.3\,
M_{\odot}$  star.  In many aspects the evolution of this star  resembles
very much that of the $9\, M_{\odot}$ single star previously  studied by
Garc\'{\i}a--Berro  et al.  (1999), the only  difference  being that the
second  dredge-up is not caused by carbon  burning, but, instead, by the
gravothermal energy release at the end of core helium burning and before
carbon is ignited off-center in the degenerate core.

Another  important  difference  between  case  1.a and  case  1.b is the
evolutionary  stage of the secondary at the time at which  reversal mass
transfer  occurs.  For the  case  1.a,  it  occurs  when  the  secondary
component  climbs  the red giant  branch,  whereas  for case 1.b it only
fills its Roche lobe when it reaches the Super--AGB phase.  However, the
uncertainties  related to stellar winds, both in mass-loss  rates and in
the efficiency of mass accretion, do not allow us to determine the exact
orbital  parameters  at the  begining of  reversal  mass  transfer  and,
ultimately,  whether merger episodes could occur.  In fact, if mass loss
from the  secondary  (as a red giant or as a Super--AGB  star)  allows a
significant  decrease in its mass with low enough mass  transfer  rates,
the steep  change in the density  profile of the mass losing star at the
border of its core could brake the spiral-in of the primary.  The system
would  then  be  able  to  survive  as  a  binary,  and  the   following
possibilities may arise:

\begin{enumerate}

\item Let us  consider  case  1.a.  If  most  of the  mass  lost  by the
      secondary  component during the reversal mass transfer  episode is
      lost by the system and the primary  accretes matter at rates lower
      than the critical value for hydrogen  burning, we might  encounter
      that the system  experiences nova outbursts.  The final outcome of
      the primary  component will still be a white dwarf but its massive
      companion will evolve to undergo a supernova  outburst and leave a
      neutron  star  as a  remnant.  The  highly  eccentric  pulsar  PSR
      B2303+46  (Van  Kerkwijk \&  Kulkarni,  1999)  could be a possible
      observational  counterpart for such a system, as the companion for
      the neutron star is a massive ($1.2\,  M_{\odot}  \lesssim  M_{\rm
      WD} \lesssim 1.4\,  M_{\odot}$)  white dwarf that might correspond
      to our primary  component  after having  accreted  some extra mass
      during the reversal mass transfer episode.

      If, on the other hand,  accretion  onto the  primary is  efficient
      enough so that this  component is able to reach the  Chandrasekhar
      mass, it will become a neutron  star after a  supernova  explosion
      (Guti\'errez  et al.  1996).  Furhermore,  if during the accretion
      phase of the primary, the  secondary  does not lose a  significant
      amount of mass, our system may help to explain the HMXB  precursor
      PSR J1740-3052 (Stairs et al., 2001), a binary system with $P_{\rm
      orb}=234$ days, composed of a neutron star plus a massive ($\grsim
      10\,  M_{\odot}$)   companion.  Similarly,  the  highly  eccentric
      pulsar PSR B1259-63  (Johnston et al., 1992) might also be another
      possible   observational   counterpart,   as  the   non-degenerate
      companion of the neutron star has a mass $\grsim 10\, M_{\odot}$.

      Further evolution of the secondary might also lead it to undergo a
      supernova  explosion and leave a second neutron star as a remnant,
      and hence a binary pulsar could also be a possible  outcome.  Such
      a system  may  remain  bound  or be  disrupted,  depending  on the
      asymmetry  of the  supernova  explosion  and on the amount of mass
      ejected  from the  system.  A possible  observational  counterpart
      could be the binary pulsar PSR B1813+16 (Taylor et al., 1976).

\item Concerning  our scenario 1.b, we have the following  possibilities
      during the reversal mass transfer  process.  If the  efficiency in
      the ejection of the common  envelope  formed during mass loss from
      the $9\,  M_{\odot}$  Super-AGB star is high, the final outcome of
      the system will be a very close double white dwarf (Maxted,  Marsh
      \& Moran 2002), possibly after having experienced a series of nova
      outbursts.  The orbital  shrinkage  allowing the components to get
      close enough to interact would be a consequence of the last common
      envelope phase.  But, if the primary component is able to grow and
      reach  the  Chandrasekhar  mass,  the  outcome  will  be a  system
      composed  of  a  neutron  star  plus  a  CO  white  dwarf,   whose
      observational  counterpart  might be PSR  J1756-5322  (Edwards  \&
      Bailes, 2001),  provided that the orbital  shrinkage due to common
      envelope  evolution  might  account for the short period  ($\simeq
      11^{\rm h}$) of this system.  In this case the eccentricity of the
      system would be small, since the common  envelope  phase for these
      systems could occur after the supernova outburst.

\end{enumerate}

Finally,  we would  like to  emphasize  that one of our  most  important
findings is that Asymptotic  Giant Branch stars could indeed be found in
binary  systems, in agreement with the  predictions of Jorissen  (1999).
The  Super-Asymptotic  Giant  Branch star is the  secondary  of a binary
system in our  scenario  1.b, and should have an evolved  (and  possibly
degenerate)  companion.  The  observational  signature  of  these  stars
should be an  anomalous  enhancement  of the  abundances  of carbon  and
oxygen and a slight underabundance of nitrogen with respect to solar.

%_______________________________________________________________________

\begin{acknowledgements}  
Part of this work was  supported  by the  Spanish  DGES  project  number
PB98--1183--C03--02,  by the MCYT grant AYA2000--1785, and by the CIRIT.
We also wish to thank J.  Jos\'e for carefully  reading the  manuscript,
and to D.  Vanbeveren, for his numerous suggestions.
\end{acknowledgements}

%_______________________________________________________________________

%_______________________________________________________________________

\end{document}